\documentclass[preprint,aps]{revtex4}

\usepackage{graphicx}
\usepackage{dcolumn}
\usepackage{bm}

\begin{document}

\preprint{APS/123-QED}

\title{Evolution of massive fields around a black hole in Ho\v{r}ava gravity}

\author{Nijo Varghese}
\author{V C Kuriakose}%
 \email{nijovarghese@cusat.ac.in, vck@cusat.ac.in}
\affiliation{%
Department of Physics, Cochin University of Science and Technology,
Cochin - 682 022, Kerala, India.
}%

\date{\today}

\begin{abstract}
We study the evolution of massive scalar field in the spacetime
geometry of Kehagias-Sfetsos(KS) black hole in deformed
Ho\v{r}ava-Lifshitz(HL) gravity by numerical analysis. We find that
the signature of HL theory is encoded in the quasinormal mode(QNM)
phase of the evolution of field. The QNM phase in the evolution
process lasts for a longer time in HL theory. QNMs involved in the
evolution of massive field are calculated and find that they have a
higher oscillation frequency and a lower damping rate than the
Schwarzschild spacetime case. We also study the relaxation of field
in the intermediate and asymptotic range and verified that behaviors
of field in these phases are independent of the HL parameter and is
identical to the Schwarzschild case.
\end{abstract}

\pacs{Valid PACS appear here}
\maketitle

\section{Introduction}
One of the major open challenges facing theoretical physics is to
accommodate general relativity(GR) in the framework of quantum field
theory. The main problem one encounters is the non-renormalizablity
of gravity at the perturbative level. Ho\v{r}ava\cite{horava}
suggested a power-counting renormalizable theory of gravity in 3+1
dimensions inspired from the Lifshitz model now dubbed as
Horava-Lifshitz gravity. It assumes a Lifshitz-like anisotropic
scaling between space and time at short distances, characterized by
a dynamical critical exponent z = 3 and thus breaking the Lorentz
invariance. While in the IR limit it flows to z = 1, retrieving the
Einstein's GR. Even though many of the fundamental questions have
not yet been answered\cite{padilla,sot} the seemingly innovative
ideas of HL theory ignited an interest in the topic and a number of
works has been reported based on it. Since the theory has the same
Newtonian and post Newtonian corrections as those of GR, systems of
strong gravity, like black holes, are needed to get observable
deviation from the standard GR. Various black hole
solutions\cite{lmp,park,bhsol1,bhsol2,bhsol3,bhsol4,bhsol5,bhsol6,bhsol7,bhsol8,bhsol9,bhsol10,bhsol11,bhsol12,bhsol13,bhsol14,bhsol15,bhsol16,bhsol17,bhsol18}
are found in HL theory of which one with asymptotically flat
Minkowski spacetime is the KS black hole given in\cite{ks} by
applying deformation in the original theory. Various aspects of KS
black hole were explored in the
past\cite{ksasp1,ksasp2,ksasp3,ksasp4,ksasp5,ksasp6,ksasp7,ksasp8,ksasp9,ksasp10,ksasp11,ksasp12,ksasp13}.

The study of response of the black hole to external perturbations is
the best way to get a clear picture of the properties of these
exotic objects. As it is well established, the evolution of
perturbations in black hole spacetimes involves three
stages\cite{kokko}. The first one is an initial response containing
the information of the particular form of the original wave field
followed by a region dominated by damped oscillation of the field
called quasinormal modes, which depends entirely on the background
black hole spacetimes. The QNMs play an important role in almost all
astrophysical processes involving black holes and the spectra of
QNMs may be different in various theories of gravity and would help
us to distinguish these theories. The late time tail stage dominates
 in the final phase of evolution.

Evolution of different fields and the associated QNMs were studied
in HL black holes in the frequency
domain\cite{ding,majhi,songbai,konoplyaqnm,wang,lin} and in time
domain\cite{nv}. All these studies on the field evolution around
black holes in HL theory has so far been restricted to the massless
fields. Evolution of massive field is a more intriguing subject of
study. The spectrum of entropy/area is discussed in the viewpoint of
QNMs of massive scalar field in\cite{setare}. It is well understood
in standard GR that massive field behaves differently from the
massless one. It was confirmed that the massive fields decay more
slowly than massless one\cite{massivescalar}. The late-time behavior
of the field that follows the QNM stage of evolution is also an
interesting topic of study since it reveals the actual physical
mechanism by which a perturbed black hole sheds its hairs. It was
first demonstrated by Price\cite{price} that in the background of
the Schwarzschild spacetime the massless neutral field at late time
dies off as $\Psi\sim t^{-(2\ell+2)}$ or $\Psi\sim t^{-(2\ell+3)}$
depending on the initial conditions and the multipole order $\ell$.
In contrast with the massless fields, massive fields have an
oscillatory inverse power-law behavior, with $\Psi\sim
t^{-(\ell+3/2)}sin(mt)$ at intermediate late times $Mm\ll mt \ll
1/(mM)^{2}$\cite{hod}. But it was shown analytically that in the
asymptotic late times ($mt \gg 1/(mM)^{2}$) another pattern of
oscillatory tail of the form $\Psi\sim t^{-(5/6)}sin(mt)$
dominates\cite{koyama} and it was numerically verified for various
space times and fields\cite{moderski,konoplyatail,burko,jing}. All
these studies argued that late time relaxation does not have any
relation to the space time parameters. Inspired from the hints that
QNMs show dependence on spacetime parameter in three dimensional
AdS\cite{birm} and dS\cite{abdalla} space times, a detailed
numerical study of relaxation process in RN space time was done in
\cite{xue}. They showed that for $Mm\ll 1$ relaxation depends only
on the field parameters, but when $Mm\gg 1$ spacetime parameters
affect the relaxation and found that for a Schwarzschild black hole
bigger the black hole mass is, the faster the perturbation decays.
The purpose of this paper is to investigate the different stages of
evolution of a massive scalar field around black hole in HL theory.

The paper is organized as follows. In Sec.\ref{sec2} we derive the
wave equation for massive scalar field around KS black hole and the
numerical method employed to study the time evolution is  explained
and in Sec.\ref{sec3} the evolution of a massless field is studied
using the method of time domain integration. The massive field
perturbation is studied in Sec.\ref{sec4} and the results are
summarized in Sec.\ref{sec5}.

\section{Scalar field around KS black hole}
\label{sec2} In \cite{ks} Kehagias and Sfetsos obtained a
spherically symmetric black hole solution by deforming HL gravity
that allows asymptotically flat spacetimes with the addition of a
term proportional to the Ricci scalar of three-geometry,
$\mu^{4}R^{(3)}$, with the cosmological constant
$\Lambda_{W}\rightarrow0$.  This will not alter the UV properties of
the theory but it does the IR ones leading to Minkowski vacuum
analogous to Schwarzschild spacetime in GR. The external
gravitational field of such a black hole has the form\cite{ks},
\begin{equation}
\ ds^{2}=-N(r)^{2}dt^{2}+f(r)^{-1}dr^{2}+r^{2}d\Omega ^{2},
\label{eq1}
\end{equation}

where $f(r)$ has the form,
\begin{equation}
N(r)^{2}=f(r)=\frac{2(r^{2}-2Mr+\alpha)}{r^{2}+2\alpha+\sqrt{r^{4}+8\alpha
Mr}}. \label{fofr}
\end{equation}

The event horizons are at,
\begin{equation}
r_{\pm}=M\pm \sqrt{M^{2}-\alpha}. \label{hori}
\end{equation}
When the HL parameter $\alpha=0$ the solution reduces to the
Schwarzschild spacetime. We will now consider the evolution of
scalar field in this spacetime. Scalar field evolution is governed
by the Klein-Gordon equation,

\begin{equation}
\frac{1}{\sqrt{-g}}\partial_{\mu}(\sqrt{-g}g^{\mu\nu}\partial_{\nu})\Phi-m^{2}\Phi=0,
\label{KGeqn}
\end{equation}

Resolving the scalar field into spherical harmonics,
$\Phi(t,r,\theta,\phi)=\frac{1}{r}\Psi_{\ell}(t,r)Y_{\ell
m}(\theta,\phi)$ and employing the tortoise coordinate defined by
$dr_{*}=\frac{1}{f}dr$, the above equation can be reduced to the
form,

\begin{equation}
\left(-\frac{\partial^{2}}{\partial
t^{2}}+\frac{\partial^{2}}{\partial
r^{2}_{*}}\right)\Psi_{\ell}(t,r)=-V_{\ell}(r)\Psi_{\ell}(t,r)=0,
\label{waveqn}
\end{equation}
where the effective potential $V_{\ell}(r)$ is given by

\begin{equation}
V_{\ell}(r)=f(r)\left [
\frac{\ell(\ell+1)}{r^{2}}+\frac{1}{r}\frac{\partial f(r)}{\partial
r}+m^{2}\right ] \label{potential}
\end{equation}

The behavior of the effective potential $V_{\ell}(r)$ is plotted in
Fig.\ref{potential}. In Fig.\ref{potential}(a) the dependence of the
potential with $\alpha$ is shown for massless and massive($m=0.3$)
fields. For a given mass, as the parameter $\alpha$ increases the
peak of the potential increases where as the asymptotic region of
potential is unaffected. It is clear from Fig.\ref{potential}(b)
that as the mass of the field increases the hight of the potential
increases and its asymptotic value raises as  $m^{2}$. The barrier
nature of potential will be spoiled for high field masses.

\begin{figure}[h]
\centering
\includegraphics[width=0.45\columnwidth]{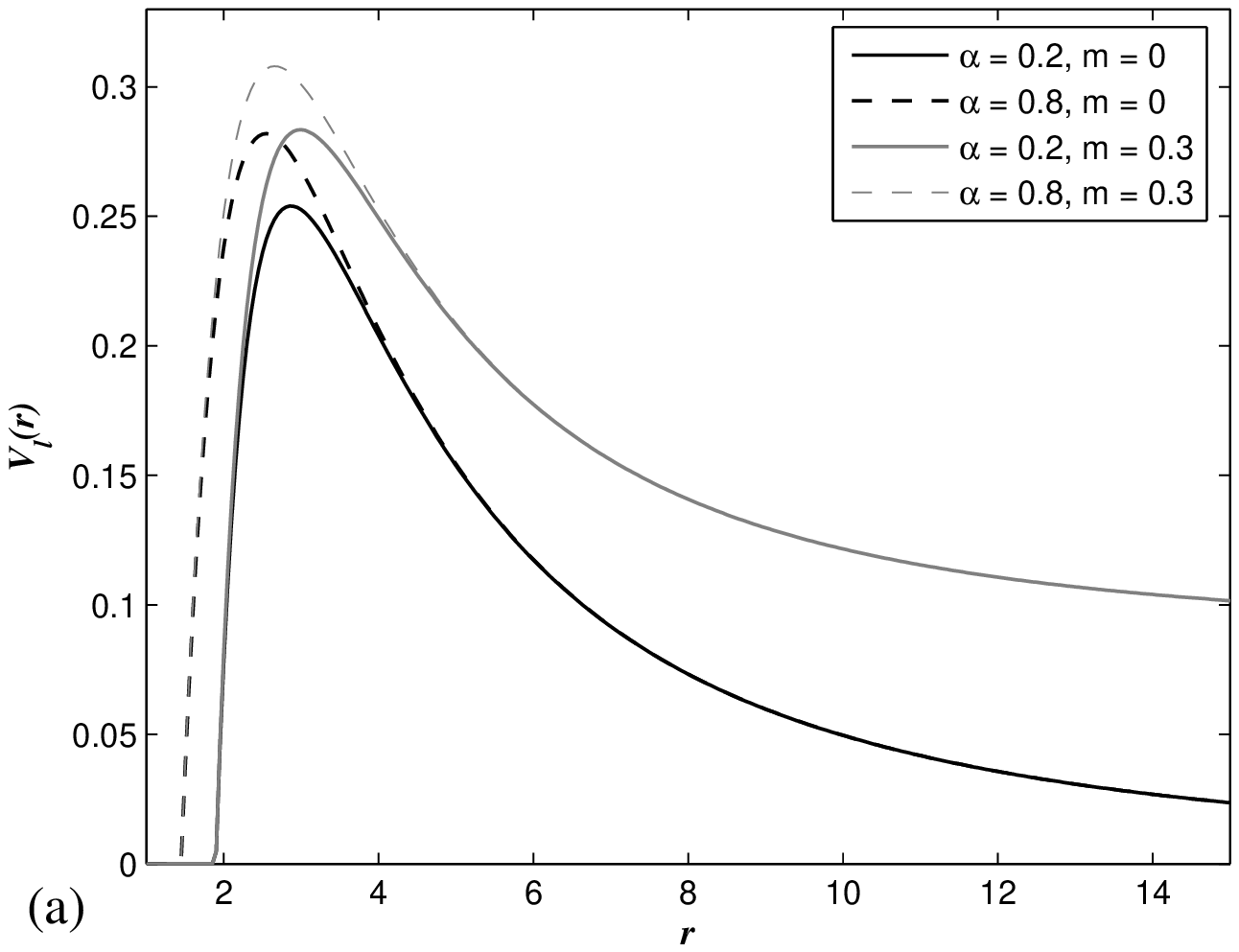}%
\hspace{0.15in}%
\includegraphics[width=0.45\columnwidth]{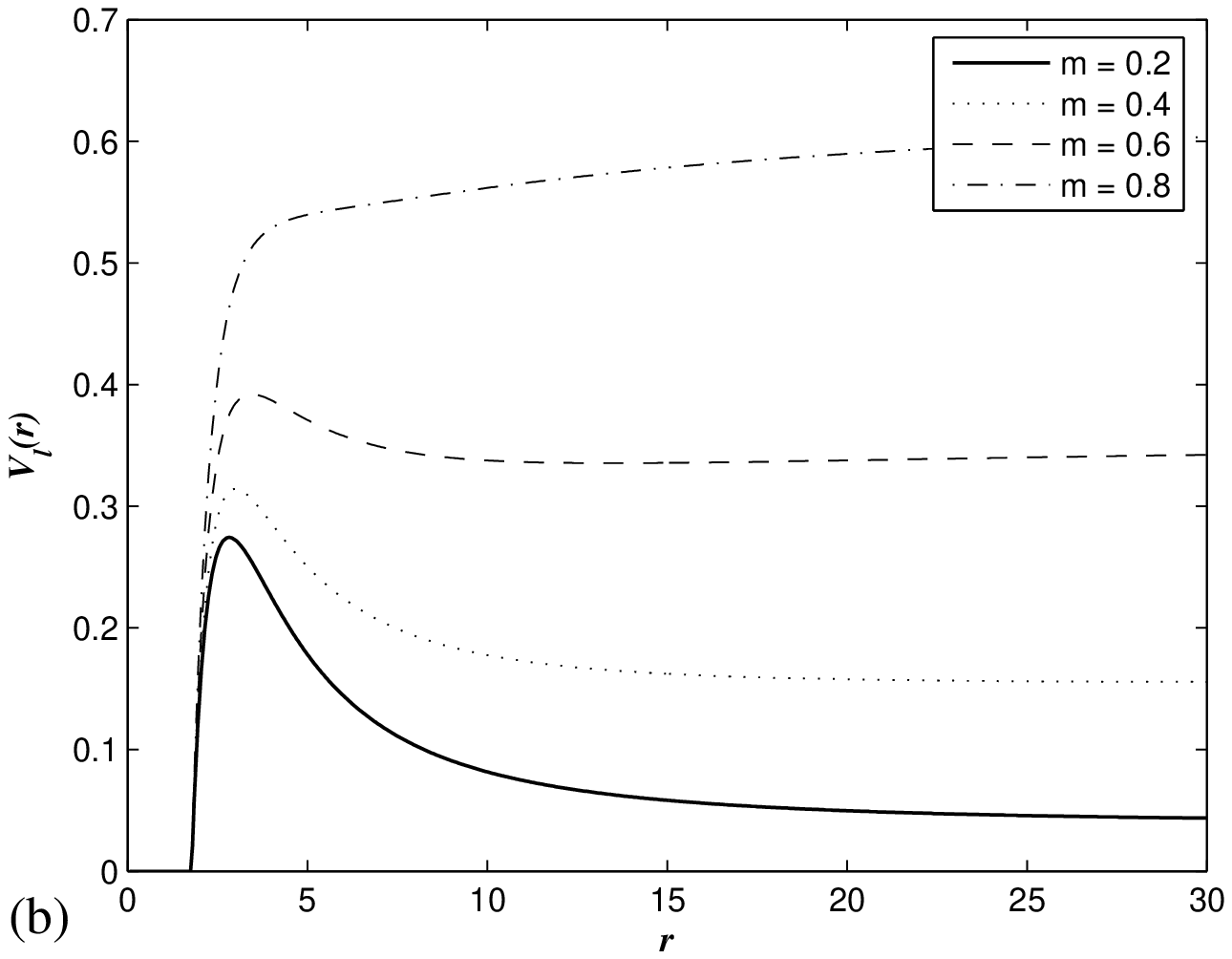}
\caption{Profiles of effective potential $V_{\ell}(r)$ for $\ell=2$.
$(a)$ $V_{\ell}(r)$ for $\alpha=0.2$(solid) and
$\alpha=0.8$(dotted). $(b)$ $V_{\ell}(r)$ with different field
masses for $\alpha=0.4$.
} %
\label{potential}
\end{figure}

The complex nature of the potential makes it difficult to obtain the
exact solutions of Eq(\ref{waveqn}) and we have to tackle the
problem by numerical methods. The wave equation Eq(\ref{waveqn}) can
be recasted in the null coordinates $u=t-r^{*}$ and $v=t+r^{*}$ as,
\begin{equation}
-4\frac{\partial^{2}}{\partial u \partial
v}\Psi(u,v)=V(u,v)\Psi(u,v). \label{waveqn2}
\end{equation}

The above equation is invariant under the resealing
\begin{equation}
 r\rightarrow ar,\quad t\rightarrow at, \quad M\rightarrow aM, \quad u\rightarrow u/a, \quad \alpha \rightarrow a^{2}\alpha
\end{equation}
for some positive constant $a$, such that $Mm$ is invariant. We
numerically integrate this equation using the difference
scheme\cite{gundlach},

\begin{equation}
\Psi_{N}=\Psi_{W}+\Psi_{E}-\Psi_{S}-\frac{h^{2}}{8}V(S)(\Psi_{W}+\Psi_{E})+O(h^{4})
\end{equation}

The integration is performed on an uniformly spaced grid with points
$N(u+h,v+h)$, $W(u+h,v)$, $E(u,v+h)$ and $S(u,v)$ forming a null
rectangle with an overall grid scale factor of $h$. Since the
late-time behavior of the wave function is found to be insensitive
to the initial data, we set $\psi(u,v=0)=0$ and a Gaussian profile
$\Psi(u=0,v)=A exp\left [-\frac{(v-v_{0})^2}{2\sigma^2} \right ]$.
In all our calculations we set the initial Gaussian with width
$\sigma=3$ centered at $v_{0}=10$. Due to the linearity of equation
one has the freedom to choose the amplitude of the initial wave
$A=1$. To proceed the integration on the above numerical scheme one
has to find the value of the potential at $r(r_{*})=r((v-u)/2)$ at
each step. For this, we numerically integrate the equation for the
tortoise coordinate using the Runge-Kutta method\cite{zhidenko} and
by cubic spline interpolation, obtained $r(r_{*})$ at each step.

\section{Evolution of massless field}
\label{sec3} 
\begin{figure}[h]
\centering
\includegraphics[width=0.45\columnwidth]{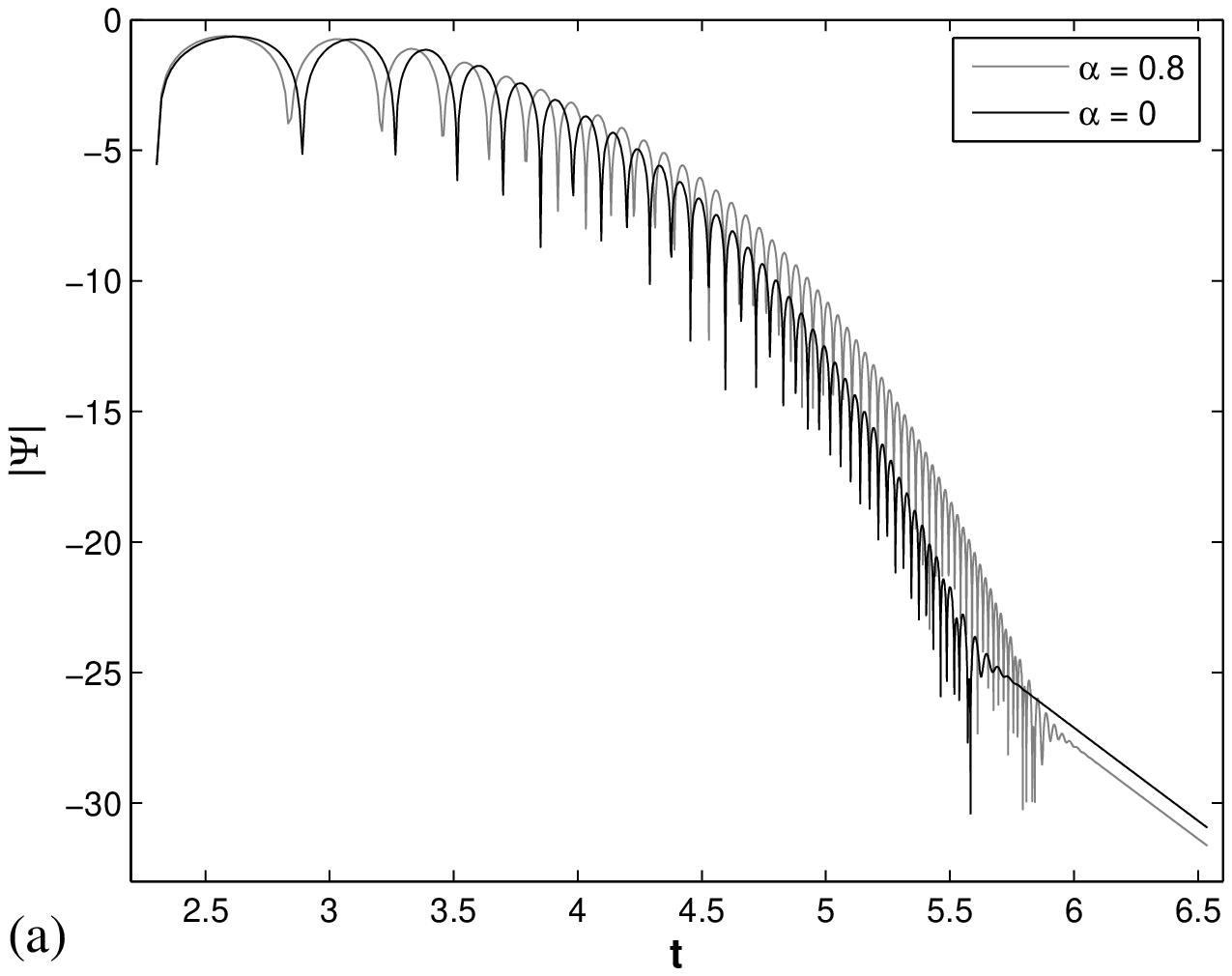}%
\hspace{0.15in}%
\includegraphics[width=0.45\columnwidth]{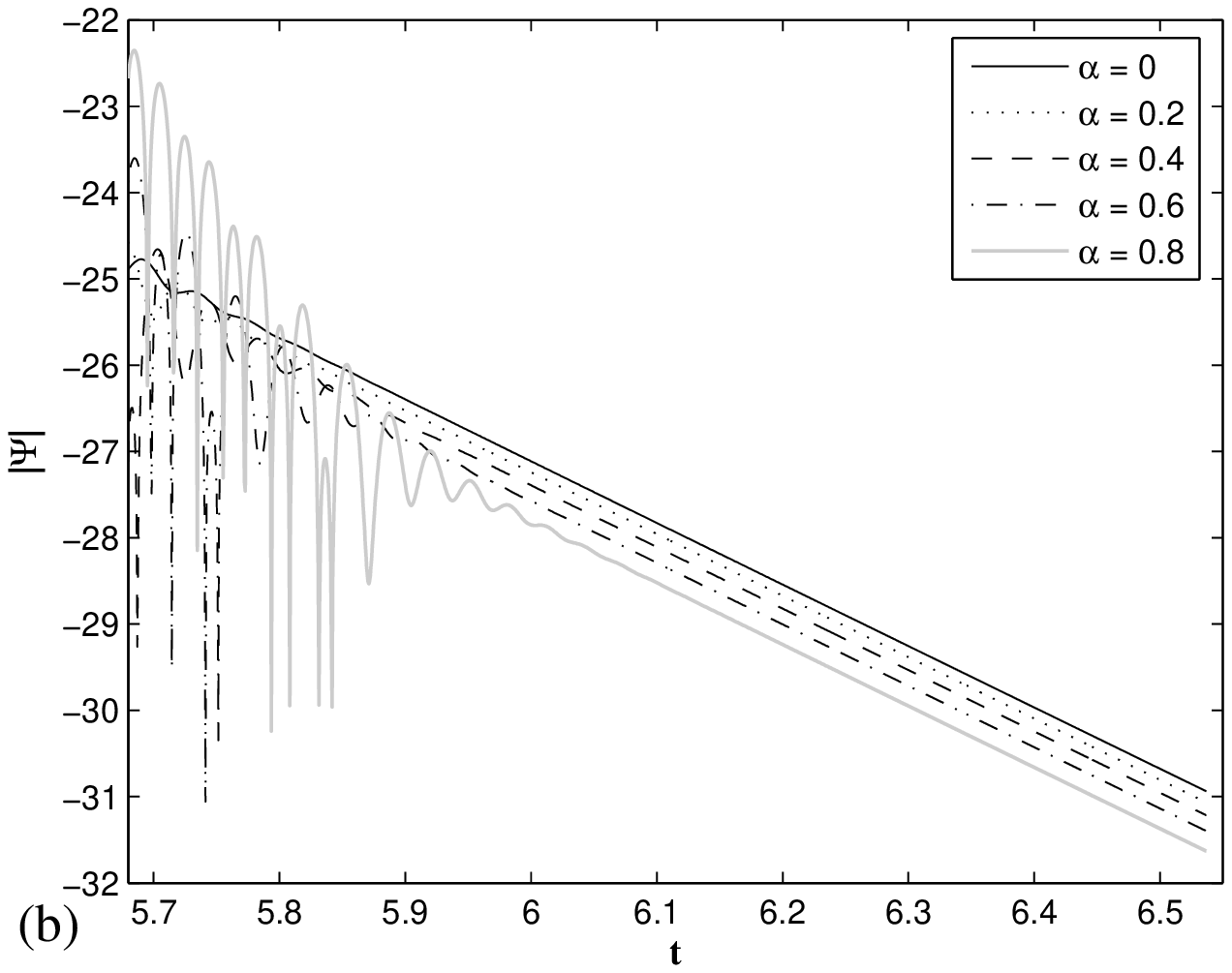}
\caption{Temporal evolution of massless scalar field in log-log
scale with $\ell=2$. $(a)$Wave function in KS spacetime with
$\alpha=0.8$ in comparison with that in Schwarzschild
spacetime($\alpha=0$). $(b)$Late time behavior of the field for
deferent values of $\alpha$.
} %
\label{mlfig1}
\end{figure}

Before studying the massive case, we first examine the
massless scalar field($m = 0$) evolution around KS black hole. In
Fig.\ref{mlfig1}(a) time evolution of a typical $\ell=2$ field in HL
theory is compared with the corresponding Schwarzschild case. We can
see that the QNM phase extends for a longer time in HL theory. The
late-time tail starts at a later time $t\approx415$ for KS black
hole with $\alpha=0.8$ whereas it is at $t\approx320$ for the
Schwarzschild case. It is also clear from the figure that the
oscillation frequency and the damping time have a higher values in
HL theory.

After the QNM stage the massless field dies off as an inverse power
of time $\Psi\sim t^{-(2\ell+3)}$. Fig.\ref{mlfig1}(b) shows the
late-time behavior of wave function for different values of
$\alpha$, with multipole index $\ell=2$. It is clear from the figure
that the late-time behavior massless field decay is independent of
$\alpha$ and is same as for the Schwarzschild black hole.

\begin{figure}[h]
\centering
\includegraphics[width=0.45\columnwidth]{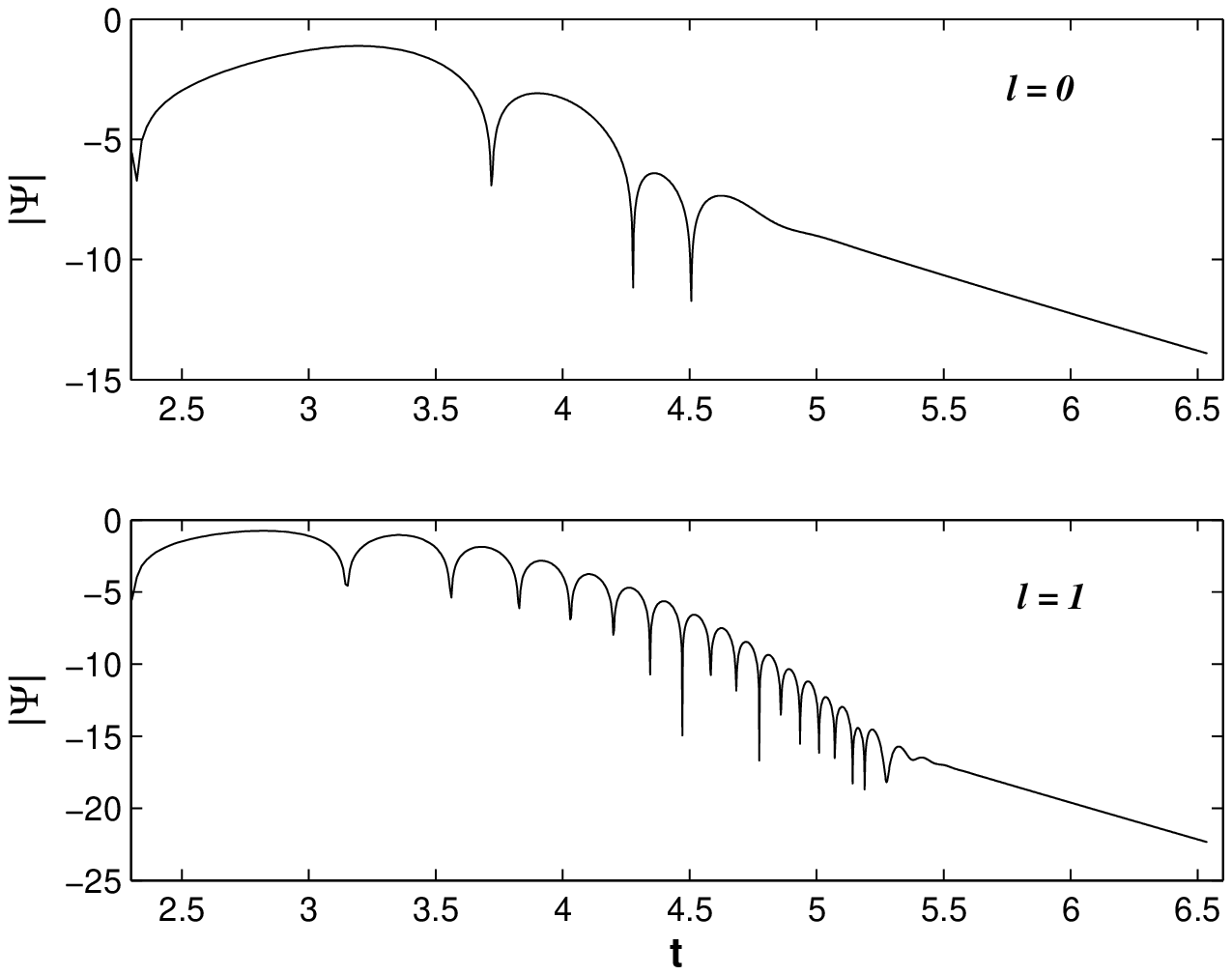}%
\hspace{0.15in}%
\includegraphics[width=0.45\columnwidth]{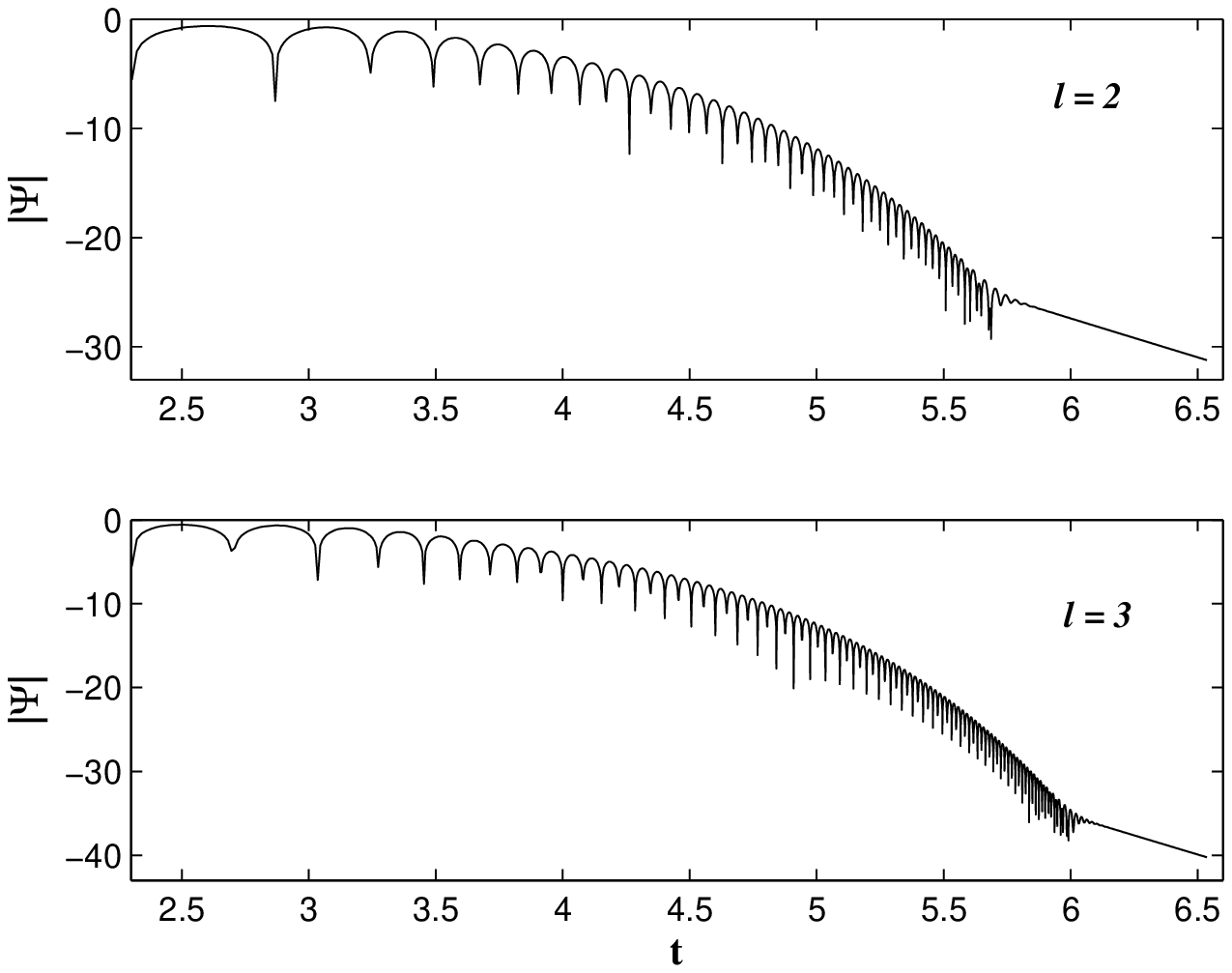}
\caption{Decay of field having different $\ell$. The late-time tail
dies off with power-law exponents $-3.14, -5.13, -7.12 $ and$ -9.13$
for $\ell=0, 1 ,2 $ and $4$ respectively.
} %
\label{mlfig2}
\end{figure}

In Fig.\ref{mlfig2}, field evolution for different multipole indices
are explored for $\alpha=0.4$. The field falls off as $\psi\sim
t^{-3.14},t^{-5.13},t^{-7.12} $ and $ t^{-9.13}$ for $\ell=0,1,2$
and $3$ respectively with the predicted powers $-3,-5,-7$ and $-9$
respectively.

\section{Evolution of massive field}
\label{sec4} 
\begin{figure}[h]
\centering
\includegraphics[width=0.45\columnwidth]{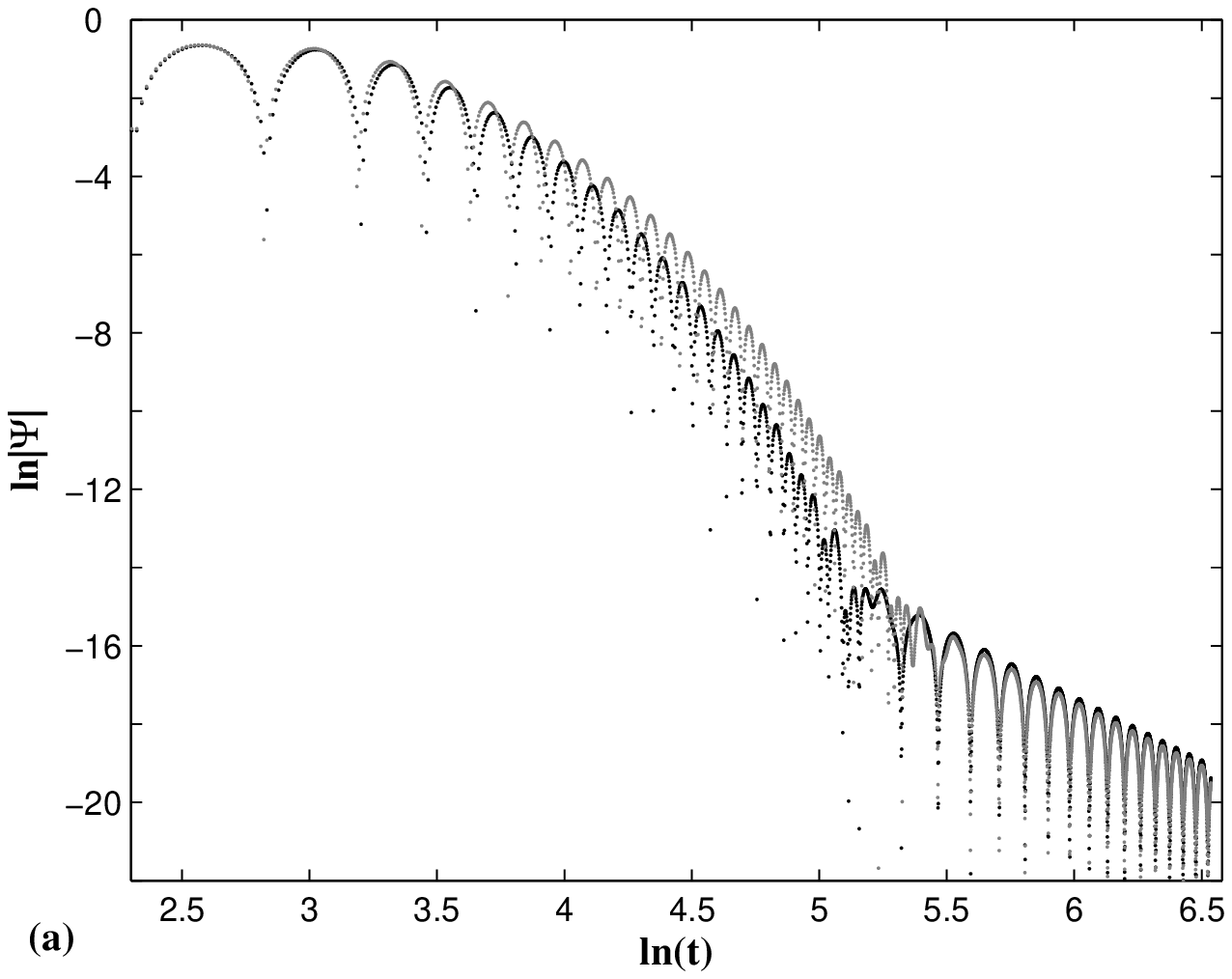}%
\hspace{0.15in}%
\includegraphics[width=0.45\columnwidth]{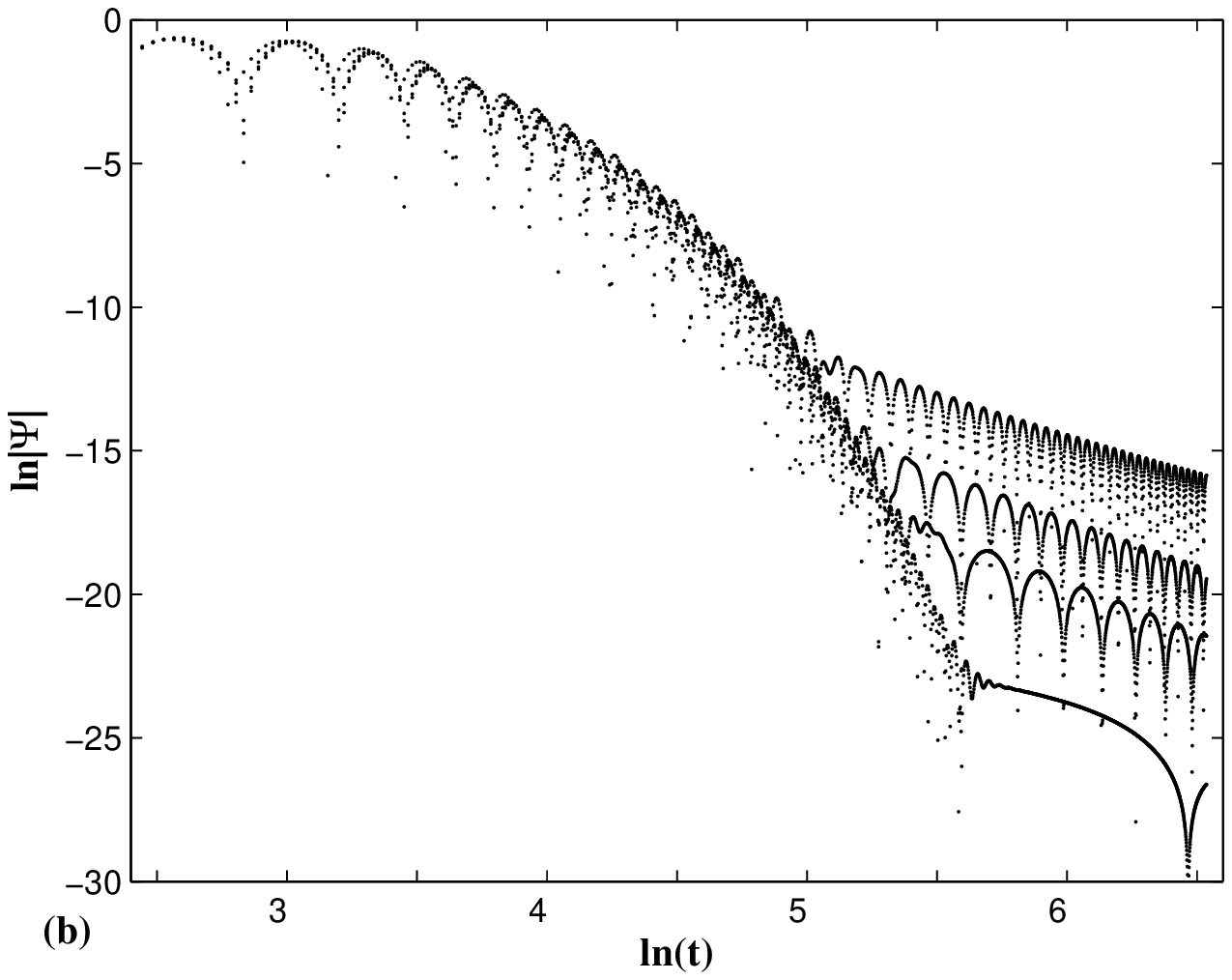}
\caption{Time evolution of massive scalar field for $\ell=2$ mode.
$(a)$Field of mass, $m=0.1$ in KS spacetime with $\alpha=0.8$(top
curve) in comparison with the corresponding case in Schwarzschild
spacetime(bottom curve). $(b)$Field with different masses for
$\alpha=0.4$, curves from bottom to top is for $m=0.01, 0.05, 0.1$
 and $0.2$.
} %
\label{mvfig1}
\end{figure}

In this section we consider the massive field evolution
in the spacetime of KS black hole. We will compare the results with
the existing results in standard GR. Here in the first series of
experiments we set the mass of the black hole, $M=1$ and choose the
the field mass $m$ such that $Mm\leq 1$. In Fig.\ref{mvfig1}(a)the
wave function for massive scalar field$(m=0.1)$ is plotted in
comparison with the corresponding Schwarzschild case for $\ell=2$.
After the prompt response in the beginning the quasinormal ringing
starts. We can see a clear difference between the two cases in the
QNM region that the QNM phase ends at a later time in HL theory and
has a lower damping rate. Fig.\ref{mvfig1}(b) shows the variation of
QNM with the field mass $m$ for $\alpha=0.4$. As the mass of the
field increases the QNM phase shrinks in time.

To see the effects of the HL parameter $\alpha$ and field mass $m$
in QNM phase, we calculate the exact values of QNMs from the
numerically integrated data in the time domain by a nonlinear
$\chi^{2}$ fitting. We also use the third order WKB
method\cite{iyerwill}, an efficient semianalytic method originally
developed by Schutz and Will\cite{schutzwill} for the lowest order.
The method is found to be accurate for low lying modes and can be
used to explore the QNM behavior of black holes for field with very
low mass. The calculated values are given in Table \ref{tab1} and
Table \ref{tab2}. QNMs in HL theory have a higher oscillation
frequency $Re(\omega)$ and a lower damping rate $|Im(\omega)|$ than
the Schwarzschild spacetime case. Also the $Re(\omega)$ and
$|Im(\omega)|$  found to be decreasing with the increase of the HL
parameter $\alpha$. WKB method gives verse results for higher field
masses since the barrier nature of potential gets spoiled at these
mass ranges. So there should be a discrepancy between the QNMs
evaluated by time domain and WKB methods at high field mass.

\begin{table}[h]
\begin{center}
\begin{tabular}{p{0.4in}p{0.7in}p{0.7in}p{0.7in}p{0.7in}}
\hline\hline
& \multicolumn{2}{c}{WKB} & \multicolumn{2}{|c}{Time domain} \\
$\alpha$ & $Re(\omega)$ & $Im(\omega)$ & \multicolumn{1}{|c}{$Re(\omega)$} & $Im(\omega)$ \\
\hline
0 & 0.48637 & -0.09572 & 0.48552 & -0.09573 \\
0.2 & 0.49388 & -0.09253 & 0.49377 & -0.09256 \\
0.4 & 0.50231 & -0.08884 & 0.50159 & -0.08888 \\
0.6 & 0.51188 & -0.08438 & 0.51142 & -0.08437 \\
0.8 & 0.52293 & -0.07868 & 0.52143 & -0.07893 \\
1 & 0.53587 & -0.07069 & 0.53569 & -0.07106 \\
\hline\hline
\end{tabular}
\caption{Fundamental($n=0$) QNM frequencies of massive($m=0.1$)
scalar field for $\ell=2$, calculated using WKB and numerical
integration data} \label{tab1}
\end{center}
\end{table}

\begin{table}[h]
\begin{center}
\begin{tabular}{p{0.4in}p{0.7in}p{0.7in}p{0.7in}p{0.7in}}
\hline\hline
& \multicolumn{2}{c}{WKB} & \multicolumn{2}{|c}{Time domain} \\
$m$ & $Re(\omega)$ & $Im(\omega)$ & \multicolumn{1}{|c}{$Re(\omega)$} & $Im(\omega)$ \\
\hline
0 & 0.49942 & -0.08969 & 0.49886 & -0.08951 \\
0.02 & 0.49953 & -0.08966 & 0.49979 & -0.08969 \\
0.05 & 0.50014 & -0.08948 & 0.49839 & -0.08948 \\
0.07 & 0.50083 & -0.08928 & 0.50156 & -0.08922 \\
0.1 & 0.50231 & -0.08883 & 0.50159 & -0.08875 \\
0.2 & 0.51101 & -0.08622 & 0.51083 & -0.08701 \\
0.3 & 0.52571 & -0.08169 & 0.52737 & -0.08029 \\
0.4 & 0.54665 & -0.07501 & 0.53933 & -0.06758 \\
0.5 & 0.57423 & -0.06568 & 0.54165 & -0.05451 \\
\hline\hline
\end{tabular}
\caption{QNM frequencies afor different mass of the scalar field
$m$. QNMs calculated using WKB and numerical integration data
($\ell=2$ and $\alpha = 0.4$)} \label{tab2}
\end{center}
\end{table}

\begin{figure}[h]
\centering
\includegraphics[width=0.45\columnwidth]{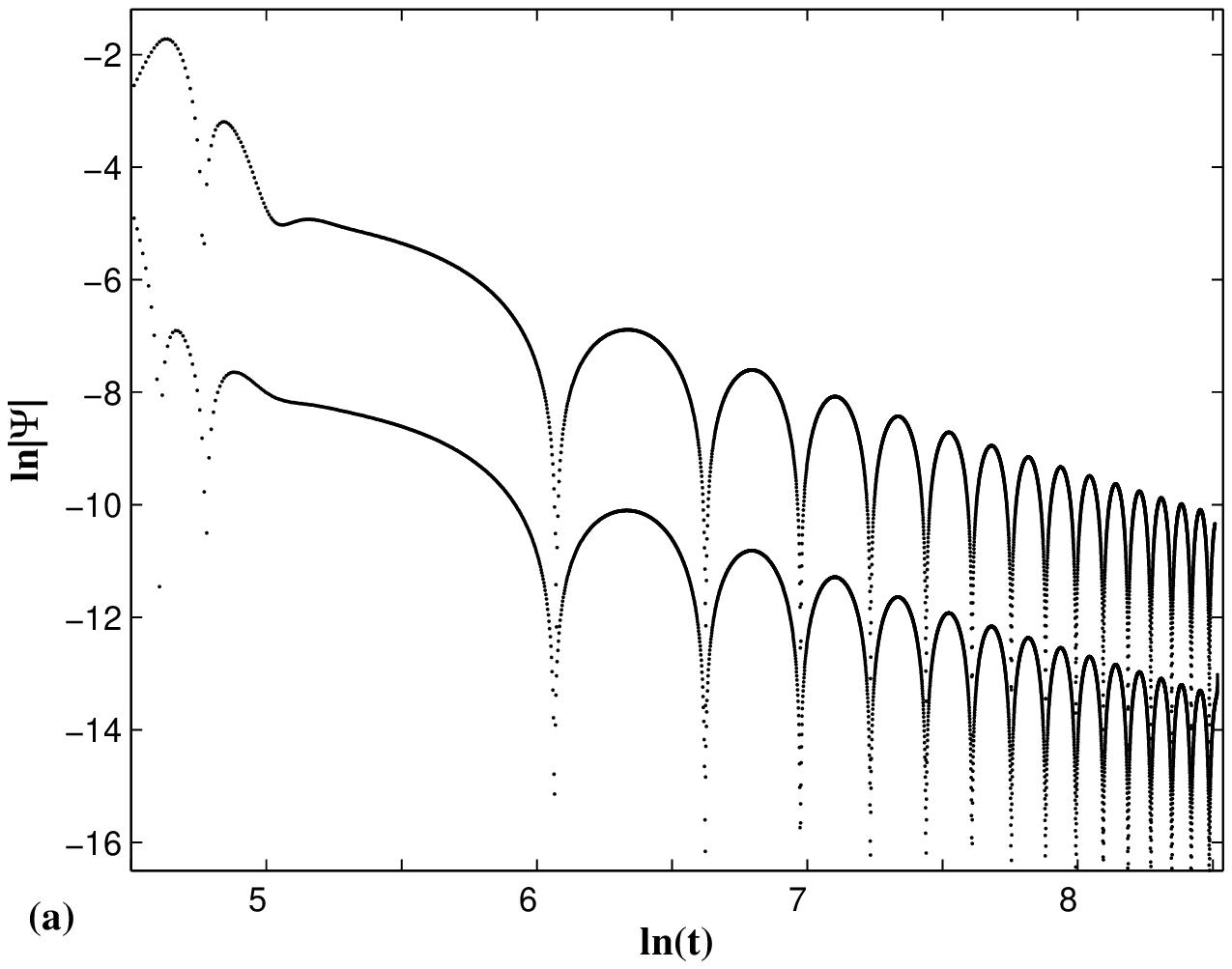}%
\hspace{0.15in}%
\includegraphics[width=0.45\columnwidth]{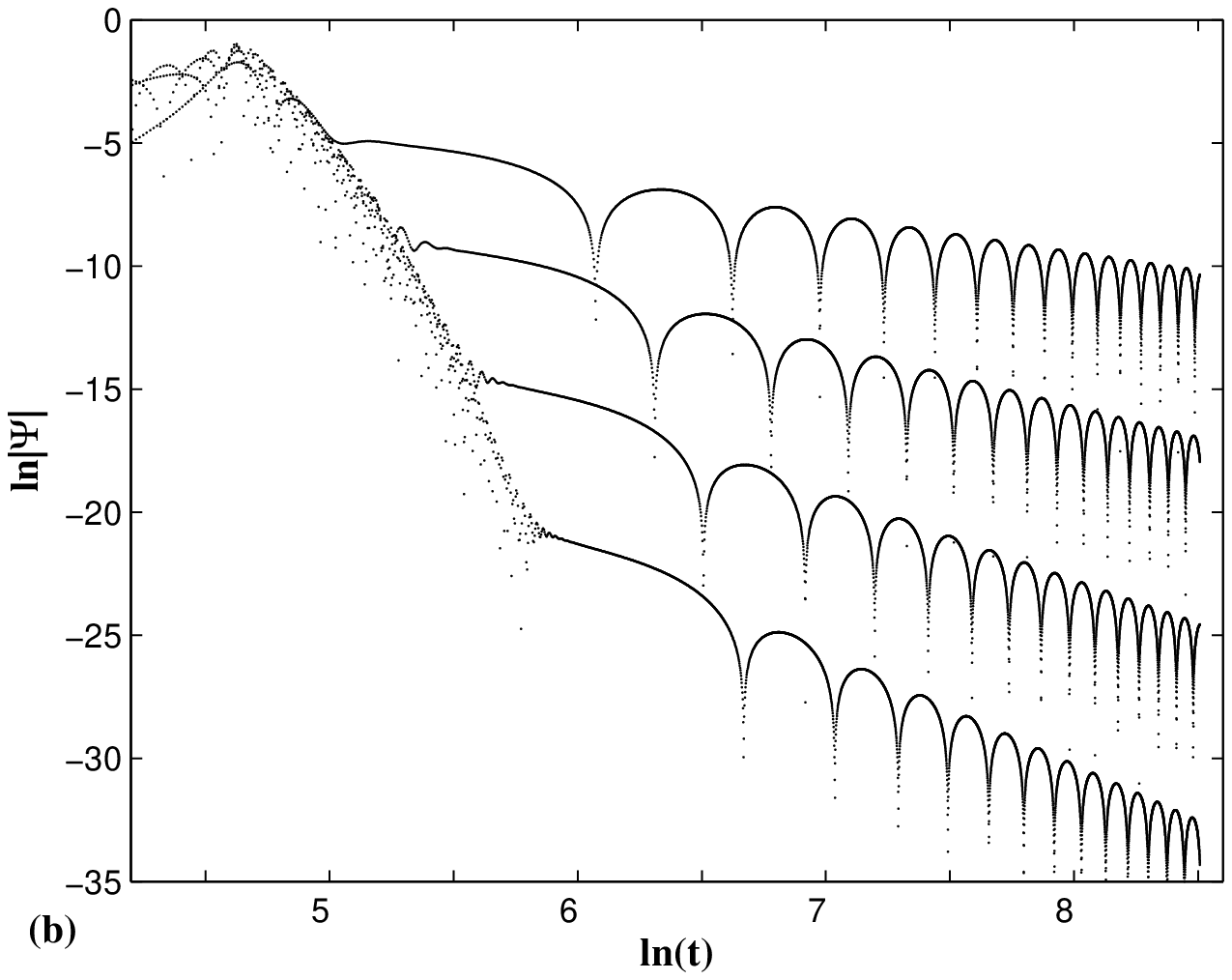}
\caption{Decay of massive field with $m=0.01$. $(a)$ Field along
outer horizon $H_{+}$(top curve) and future time like infinity
$i_{+}$ (bottom curve). The amplitude of field decay as power law
with exponent $-1.49$ and period of oscillation $T=314.8$$(b)$ Field
along $i_{+}$ for different $\ell$. The field dies off as power-law
with exponents $-1.49, -2.5, -3.51$ and $-4.51$ for $\ell=0,1,2,$
and 3 respectively. The period of oscillation $T=314.7\pm 0.03$.
} %
\label{mvfig2}
\end{figure}

The QNM phase is followed by the phase of late time tail behavior of
the field decay. First we consider the late time tails in
intermediate range, $Mm\ll mt \ll (mM)^{2}$. The decay of field
along black hole outer horizon $H_{+}$(approximated by the null
surface $u_{max}=0.5\times10^{4}$) and future time like infinity
$i_{+}$ (approximated on fixed radius $r_{*}=50M$) are evaluated
with the initial field profile parameters $v_{0}=50$ and
$\sigma^{2}=2$. The result for $\alpha=0.4$ with $m=0.01$ and
$\ell=0$ is shown in Fig.\ref{mvfig2}(a). After the QNM phase the
amplitude of the oscillatory field decays as a power law with
exponent -1.49 along $I_{+}$ and $H_{+}$ with a period of
oscillation $T=314.8$. We have analyzed for different values of HL
parameter $\alpha$ and found that the tail behavior is almost
independent of $\alpha$ and the decay is identical to the
Schwarzschild case according to the form,

\begin{equation}
\Psi\sim t^{-(\ell+3/2)}sin(mt) \label{tail}
\end{equation}

This can be expected since the late-time tail is originated by the
backscattering by the effective potential in the asymptotic region
and as it is clear from Fig.\ref{potential}, that the parameter
$\alpha$ has no effect on this asymptotic region of the potential
but can only change its shape near the peek.

The field decay for different multipole indices is plotted in
Fig.\ref{mvfig2}(b) with $m=0.01$. The decaying tail is found to
have a period of oscillations $T=314.7\pm 0.03$ and the power-law
exponents $-1.49, -2.5, -3.51$ and $-4.51$ for $\ell=0,1,2,$ and 3
respectively. An excellent agrement with the decay rate of the form
$t^{-(\ell+3/2)}$ can be seen. We find that the frequency and
damping rate of the oscillatory tails are independent of the
parameter $\alpha$ and follow the Schwarzschild case. Thus
Eq(\ref{tail}) will be the form of the intermediate late-time
behavior of massive field decay in HL theory also.

\begin{figure}[h]
\centering
\includegraphics[width=0.45\columnwidth]{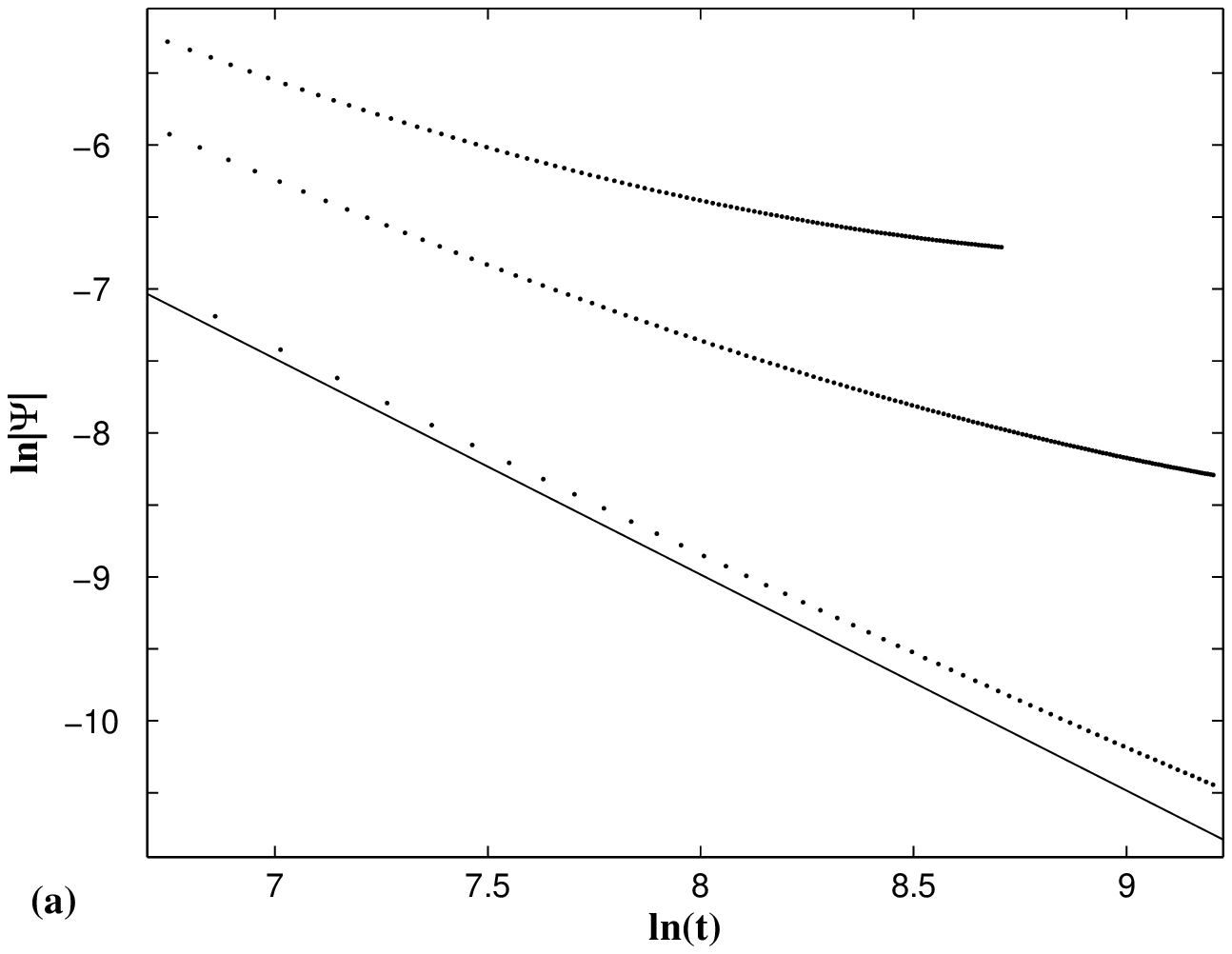}%
\hspace{0.15in}%
\includegraphics[width=0.45\columnwidth]{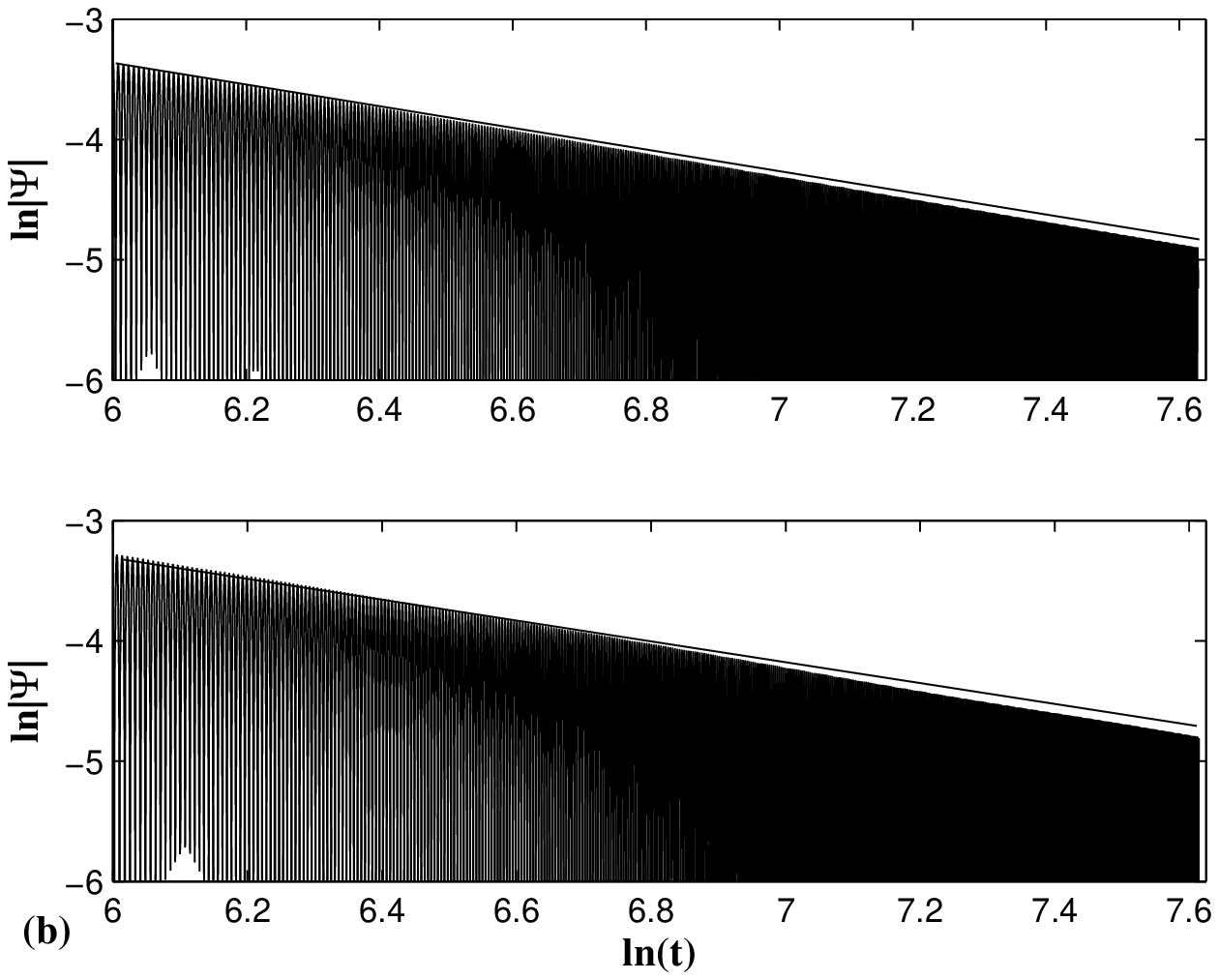}
\caption{Late time behavior of field for $\alpha = 0.4$. $(a)$Maxima
of oscillation for $\ell=0$, from bottom to top $m = 0.02, 0.05,
0.07$. Solid line has a slop $-1.5$. $(b)$Asymptotic regime of field
decay with $m=1$ for $\ell=0$(top) and $\ell=2$(bottom). Straight
lines have a slop $-5/6$.
} %
\label{mvfig3}
\end{figure}

But in the asymptotic late time the decay of field does not follow
pattern given by Eq(\ref{tail}). Another pattern of oscillatory tail
dominates in the asymptotic late times $mt\gg 1/(Mm)^{2}$. This can
be seen from Fig.\ref{mvfig2}(a), where the maxima of oscillations
are shown for $\alpha=0.4, \ell=0$ and $m=0.02, 0.05, 0.07$. The
deviation from straight line with slop $-1.5$ is visible in the
asymptotic late time. When $mt\gg 1/(Mm)^{2}$ smaller the $Mm$
 the later the asymptotic tail starts. In Fig.\ref{mvfig2}(b)
asymptotic region is shown for $m = 1$ and $\ell=0$ and $2$. We find
that the asymptotic tail is independent of the multipole order and
follows the $\Psi\sim t^{-5/6}sin(mt)$ form as in the standard GR.
We have analyzed the asymptotic tail for different values of
$\alpha$ and find that this regime is also independent of the HL
parameter $\alpha$.

\begin{figure}[h]
\centering
\includegraphics[width=0.45\columnwidth]{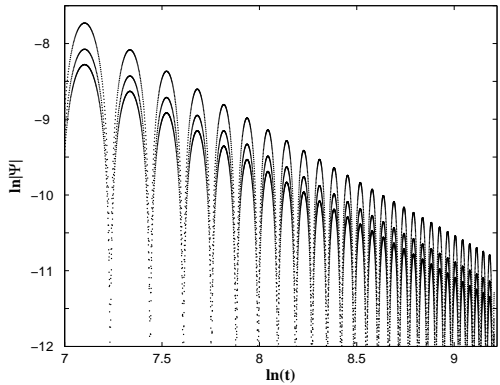}
\caption{Decay of field with different black hole mass for $m=0.01$,
$\alpha=0.4$ and $\ell=0$. For $M=0.5,1$ and $1.5$(curves from top
to bottom) the decay rates of field are $-1.458,-1.459$ and $-1.462$
respectively.
} %
\label{mvfig4}
\end{figure}

Now we study the role of the black hole parameter on the decay of
field for $Mm\ll1$. The numerical results for different black hole
mass with field mass $m=0.01$ are shown in Fig.\ref{mvfig4}. The
field falls off with power law exponents $-1.458,-1.459$ and
$M=0.5,1$ and $1.5$ respectively and we can conclude that for
$Mm\ll1$ the decay rate is independent of black hole mass. We are
not going for the  $Mm\gg 1$ case of computationally expensive part
of study. For $Mm\gg 1$ range we have good reasons to believe that
the relaxation will show similar results of the Schwarzschild case
as shown in\cite{xue}, since we have already shown that the
parameter $\alpha$ has no effect on the asymptotic region of the
potential and the late time relaxation is independent of the HL
parameter.

\section{Summary}
\label{sec5} We have studied the time evolution of massive scalar
field in the spacetime geometry of KS black hole in deformed HL
gravity by time domain integration. We find a noticeable deviation
of the evolution behavior of massive field in the ringdown region
from the standard Schwarzschild black hole. QNMs involved in the
evolution of massive field are calculated from the time domain
integration data and third order WKB method. Massive QNMs in HL
theory have a higher oscillation frequency and a lower damping rate
than the Schwarzschild spacetime case. The ringdown frequency and
damping rate also decrease with the increase of the HL parameter
$\alpha$. As the field mass increases the QNM phase squeezes to a
smaller time interval. However the late time evolution of massive
field fails to show any distinction from the Schwarzschild case. In
the intermediate range the field decays as $t^{-(\ell+3/2)}sin(mt)$,
but in the asymptotic late time the decay is dominated by
$t^{-5/6}sin(mt)$ tail. In the range $mt\gg 1/(Mm)^{2}$, smaller the
value $Mm$ the later the asymptotic tail starts. We also shown that
for $Mm\ll1$ the relaxation is independent of space time parameters.
Since the HL parameter $\alpha$ has no effect on the asymptotic
region of the potential the late time relaxation is independent of
the HL parameter.

\section*{Acknowledgements} NV wishes to thank UGC, New Delhi for financial support under RFSMS
scheme. VCK is thankful to CSIR, New Delhi for financial support
under Emeritus Scientistship scheme and wishes to acknowledge
Associateship of IUCAA, Pune, India.

\end{document}